\documentclass[comment, prl,twocolumn,nofootinbib, preprintnumbers, superscriptaddress]{revtex4}

\usepackage{amsmath,amssymb,bm,slashed,braket}
\usepackage{graphicx}
\usepackage{epstopdf}
\usepackage{float}
\usepackage[colorlinks=true,
            linkcolor=blue,
            urlcolor=blue,
            citecolor=green,          
            bookmarks=true,
            bookmarksnumbered=true,
            breaklinks=true,
            pdfpagemode=Fullscreen,
            pdfstartview=FitBH]{hyperref}

\usepackage[normalem]{ulem}
\usepackage{color}

\definecolor{Orange}{cmyk}{0,0.61,0.87,0}
\definecolor{JungleGreen}{cmyk}{0.99,0,0.52,0}
\definecolor{OliveGreen}{cmyk}{0.64,0,0.95,0.40}
\definecolor{Brown}{cmyk}{0,0.81,1,0.60}
\definecolor{RoyalBlue}{cmyk}{0.71,0.53,0,0.12}
\definecolor{Gray}{cmyk}{0,0,0,0.40}
\definecolor{LightPink}{cmyk}{0.0,0.25,0,0}
\definecolor{LLightPink}{cmyk}{0.0,0.10,0,0}
\definecolor{LightBlue}{cmyk}{0.25,0,0,0}
\definecolor{LightGray}{cmyk}{0,0,0,0.2}

\usepackage{xcolor}
\definecolor{gesfpurple}{rgb}{0.47,0.19,0.42}

\definecolor{gesflanse}{rgb}{0.00,0.50,0.50}

\definecolor{gesfblue}{rgb}{0.08,0.42,0.76}

\definecolor{gesfred}{rgb}{1,0,0}

\definecolor{gesfwhite}{rgb}{1,1,1}

\definecolor{gesfblack}{rgb}{0,0,0}

\newcommand{\geqn}[1]{Eq.\,\hypersetup{linkcolor=blue}(\ref{#1})\hypersetup{linkcolor=blue}}

\newcommand{\gtab}[1]{{\hypersetup{linkcolor=gesflanse}Tab.\,\ref{#1}\hypersetup{linkcolor=blue}}}

\graphicspath{{figs/}}

\begin{document}

\title{Verifying the Resonance Schemes of Unstable Particles at Lepton Colliders}

\author{Shao-Feng Ge}
\email{gesf@sjtu.edu.cn}
\affiliation{Tsung-Dao Lee Institute \& School of Physics and Astronomy, Shanghai Jiao Tong University, China}
\affiliation{Key Laboratory for Particle Astrophysics and Cosmology (MOE) \& Shanghai Key Laboratory for Particle Physics and Cosmology, Shanghai Jiao Tong University, Shanghai 200240, China}

\author{Ui Min}
\email{ui.min@sjtu.edu.cn}
\affiliation{Tsung-Dao Lee Institute \& School of Physics and Astronomy, Shanghai Jiao Tong University, China}

\author{Zhuoni Qian}
\email{zhuoniqian@hznu.edu.cn}
\affiliation{School of Physics, Hangzhou Normal University, Hangzhou, Zhejiang 311121, China}

\begin{abstract}
We propose practical ways of differentiating the
various (Breit-Wigner, theoretical, and
energy-dependent) resonance schemes of unstable
particles at lepton colliders.
First, the energy-dependent scheme can be distinguished
from the other two by fitting the $Z$ lineshape scan and
forward-backward asymmetries at LEP and future
lepton colliders with the $Z$ mass
$m_Z$, decay width $\Gamma_Z$, and coupling strength
as fitting parameters.
Although the Breit-Wigner and theoretical schemes
work equally well, the scheme conversion requires
the decay width $\Gamma_Z$ to scale inversely with $m_Z$
rather than the usual linear dependence from theoretical
calculation. These contradicting behaviors can be
used to distinguish the Breit-Wigner and theoretical
schemes by the precision $Z$ measurements with
single parameter ($m_Z$) fit at future lepton
colliders. For the $WW$ threshold scan, its combination
with the precise Fermi constant provides another way
of distinguishing
the Breit-Wigner and theoretical schemes.
\end{abstract}

\maketitle

{\bf Introduction}
--
The description of unstable particles is an intriguing part
of quantum and field theories.
A commonly adopted parametrization is the so-called
Breit-Wigner resonance \cite{Breit:1936zzb},
\begin{align}
  \mathcal D_{\rm BW}
\sim
  \frac 1 {p^2 - m^2_{\rm BW} + i m_{\rm BW} \Gamma_{\rm BW}}
\label{eq:BW}
\end{align}
in natural units ($\hbar = c = 1$),
whose magnitude squared $1 / [(p^2 - m^2_{\rm BW})^2 + m^2_{\rm BW} \Gamma^2_{\rm BW}]$
follows a resonance shape with pole at exactly the
mass, $p^2 = m^2_{\rm BW}$.
The Breit-Wigner (BW) scheme give a very
intuitive understanding of the pole mass
$m_{\rm BW}$. Since the propagator can
enter concrete processes in various ways, we refrain
from replacing $p^2$ by the Mandelstam variable
$s$ that is usually for the
$s$-channel processes to make it general.
A review of more details can be found
in \cite{Bohm:2004zi, Giacosa:2021mbz}.

A physical particle should
have definite dispersion relation, $E = \sqrt{\bm p^2 + m^2}$
where $(E, \bm p)$ is the 4-momentum and $m$ is mass.
With a real energy $E$, the particle state evolves into the infinite
future with a time-dependent evolution phase $e^{- i E t}$.
So a real energy $E$ associated with time evolution intrinsically
describes a stable particle. On the other hand, an unstable particle
develops an imaginary part $- \frac i 2 \Gamma$ that is attached
to the real energy, $E \rightarrow E - \frac i 2 \Gamma$.
Then the particle wave function
$\sim e^{- i E t - \frac 1 2 \Gamma t}$ keeps decreasing
with time and the corresponding decay width (lifetime)
is $\Gamma$ ($1/\Gamma$).

It is much more intuitive to define the exact value of
decay width in the rest frame. In other words,
the physical pole should actually happen at
$E = m - \frac i 2 \Gamma \equiv \mu$,
which is consistent with the linear approximation
for the non-relativistic limit
\cite{Dalitz:1961dx,Dalitz:1963xk}.
Considering the fact that a relativistic dispersion relation
should be Lorentz invariant,
the propagator contains $p^2$
in the denominator \cite{Willenbrock:1991hu, Willenbrock:2022smq},
\begin{align}
  \mathcal D_{\rm th}
\sim
  \frac 1 {p^2 - \mu^2}
=
  \frac 1 {p^2 - m^2_{\rm th} + \frac 1 4 \Gamma^2_{\rm th} + i m_{\rm th} \Gamma_{\rm th}}.
\label{eq:th}
\end{align}
The pole $E = \sqrt{\bm p^2 + \mu^2}$ then happens at
$\mu$ in the rest frame with vanishing $\bm p$. Since this
scheme comes from theoretical arguments and
has the benefit of being gauge independent
\cite{Sirlin:1991fd, Papavassiliou:1995fq}, we call it the theoretical
(th) scheme.

Comparing with the Breit-Wigner resonance in \geqn{eq:BW},
the theoretical scheme has an extra term $\frac 1 4 \Gamma^2$.
For a narrow resonance, $\Gamma \ll m$, the Breit-Wigner
scheme is a good enough approximation. By solving the equations, $m^2_{\rm BW} = m^2_{\rm th} - \frac{1}{4} \Gamma^2_{\rm th}$ and $m_{\rm BW} \Gamma_{\rm BW} = m_{\rm th} \Gamma_{\rm th}$, however,
the $Z$ mass can shift by $m_{\rm th} - m_{\rm BW} = 8.5$\,MeV
which seems to be already larger than the current
precision $\Delta m_Z = 2.1$\,MeV \cite{Workman:2022ynf}.

Another commonly used scheme for the $Z$ pole scan is
the energy-dependent scheme
\cite{Berends:1987bg, Bardin:1988xt, Bardin:1989di},
\begin{align}
  \mathcal D_{\rm ED}
\sim
  \frac 1 {p^2 - m^2_{\rm ED} + i \frac {p^2 \Gamma_{\rm ED}}{m_{\rm ED}}},
\label{eq:ED}
\end{align}
with the $Z$ decay width
$\Gamma(p^2) \sim \frac {p^2} {m^2_{\rm ED}} \Gamma_{\rm ED}$
linearly scales with $p^2$. Besides mass $m$ in the
prefactor, the
scattering matrix and phase space combination of
the decay width 
$\Gamma \equiv \frac 1 {2 m} \int |\mathcal M|^2 d \Omega$
is Lorentz invariant and hence a function of the
virtuality $p^2$. So it is understandable,
$\Gamma(p^2) \propto p^2 / m$ scales linearly with
virtuality $p^2$ above the decay threshold.
The energy-dependent scheme has
been used in the LEP data analysis
\cite{WorkingGrouponLEPEnergy:1993sza,ALEPH:2005ab} and
Zfitter \cite{Bardin:1999yd}.

Since the propagator schemes would affect the
interpretation of the experimental data almost everywhere,
especially for the precision test of new physics beyond
the Standard Model (SM) of particle physics, it is of
great importance to testify them with real
data which is still an open question.
We would thoroughly explore the phenomenological consequences
with the $Z$ lineshape and $WW$ threshold
scans at lepton colliders.
The precision measurement of Fermi constant $G_F$
can also provide an input for the $W$ mass.
For convenience, we take the commonly-used Breit-Wigner
scheme as anchor in our following discussions.

\vspace{2mm}

{\bf Breit-Wigner vs Energy-Dependent Schemes}
--
The comparison between the Breit-Wigner scheme in \geqn{eq:BW}
and its energy-dependent counterpart in \geqn{eq:ED} should
be done at the amplitude squared level. This is because
the energy-dependent scheme leads to,
\begin{align}
  |\mathcal D_{\rm ED}|^2
\sim
  \frac {\mathcal Z} {(p^2 - \mathcal Z m^2_{\rm ED})^2 + \mathcal Z^2 m^2_{\rm ED} \Gamma^2_{\rm ED}},
\end{align}
where $\mathcal Z \equiv 1 / (1 + \Gamma^2_{\rm ED} / m^2_{\rm ED})$
with the $p^2$ correction from the decay width term.
Comparing with the traditional Breit-Wigner scheme
$|\mathcal D_{\rm BW}|^2 \sim 1 / [(p^2 - m^2_{\rm BW})^2 + m^2_{\rm BW} \Gamma^2_{\rm BW}]$,
the peak location and decay width shifts as,
\begin{align}
  m_{\rm BW} = \sqrt{\mathcal Z} m_{\rm ED},
\quad \mbox{and} \quad
  \Gamma_{\rm BW} = \sqrt{\mathcal Z} \Gamma_{\rm ED}.
\label{eq:Zadjust}
\end{align}
In addition, the overall
$\mathcal Z$ factor can be absorbed into the field redefinition.
Since each propagator connects two vertices or two fields,
each side should receive a scaling factor of $\sqrt{\mathcal Z}$.
Take the fermion gauge coupling vertex $V \bar f f$ with
$V = Z$ or $W$ as example, the coupling strength $g$ should
be rescaled/renormalized as
$g_{\rm BW} = \mathcal Z^{1/4} g_{\rm ED}$.

The Breit-Wigner and energy-dependent schemes agree
quite well with each other for the narrow width approximation
($\Gamma \ll m$). Taking the weak gauge bosons for
illustration, the width is just 2.7\% of the mass for both $Z$
($m_Z = 91.1876 \pm 0.0021$\,GeV vs $\Gamma_Z = 2.4955 \pm 0.0023$\,GeV
at LEP 1 \cite{Workman:2022ynf,ALEPH:2005ab})
and $W$
($m_W = 80.376 \pm 0.033$\,GeV vs $\Gamma_W = 2.195 \pm 0.83$\,GeV
at LEP 2 \cite{Workman:2022ynf,ALEPH:2013dgf}).
Consequently, the relative difference $\Gamma^2 / 2 m^2$
between the two schemes is only at the level of $3.7 \times 10^{-4}$
for the $Z$ and $W$ bosons. This corresponds to shifts of 34\,MeV
and 30\,MeV in the $Z$ and $W$ masses, respectively,
and correspondingly a shift of 0.93\,MeV and 0.82\,MeV
in their decay widths. Although such correction seems
quite tiny, the sensitivity at lepton colliders can reach the precision. \footnote{The recent measurements
of $m_W$ and $\Gamma_W$ at the LHC achieved a precision
of $\Delta m_W = 15.9$\,MeV and $\Delta \Gamma_W = 47$\,MeV
from the ATLAS Collaboration \cite{ATLAS:2024erm}.
Similarly, the CMS Collaboration reached $\Delta m_W = 9.9$\,MeV
\cite{CMS:2024nau} while $\Delta \Gamma_W$ was not explicitly
reported. To distinguish between the two resonance schemes,
high precision in both the mass and decay width measurements
is essential. Despite having achieved sufficient precision
in mass measurements, it is still challenging for hadron colliders
to differentiate between the resonance schemes.} For the $Z$ boson mass, the current precision
that mainly comes from LEP is 2.1\,MeV while it
can further shrink to
0.1\,MeV at CEPC \cite{CEPCPhysicsStudyGroup:2022uwl}
and FCC-ee \cite{Blondel:2021ema}. It seems possible
and is of great
importance to use the in-situ
$Z$ lineshape scan to test the resonant schemes
for an unambiguous conclusion.

However, since the propagator amplitude squares
$|\mathcal D_{\rm BW}|^2$ and
$|\mathcal D_{\rm ED}|^2$ can be made equivalent,
there is no essential difference between
these two schemes. If only a single $Z$ propagator is
involved, the resonance scan cannot tell the difference
between the Breit-Wigner and energy-dependent schemes.
Although the extracted $Z$ mass from the same
data but with different schemes can differ by
34\,MeV, differentiating them needs extra justification.

Fortunately, the $Z$ resonance at lepton colliders
involves not just the $Z$ propagator but also the
virtual photon ($\gamma$) mediation. Being massless,
the photon is stable with vanishing decay width.
In other words, the issue of resonance scheme appears
only for the $Z$ propagator. But the two contributions
$\mathcal M_Z$ and $\mathcal M_\gamma$ can interfere
with each other to give an interference term,
$\mathcal M_Z \mathcal M^*_\gamma$,
\begin{subequations}
\begin{align}
  (\mathcal M_Z \mathcal M^*_\gamma)_{\rm BW}
& \propto
  \frac {g^2_{Z, \rm BW} e^2 (s - m^2_{\rm BW})}
        {\left[ (s - m^2_{\rm BW})^2 + m^2_{\rm BW} \Gamma^2_{\rm BW} \right] s},
\\
  (\mathcal M_Z \mathcal M^*_\gamma)_{\rm ED}
& \propto
  \frac {\mathcal Z g^2_{Z, \rm ED} e^2 (s - m^2_{\rm ED})}
        {\left[ (s - \mathcal Z m^2_{\rm ED})^2 + \mathcal Z^2 m^2_{\rm ED} \Gamma^2_{\rm ED} \right] s},
\end{align}
\label{eq:ZA}
\end{subequations}
where $g_Z$ and $e$ stands for the weak and electric
couplings. Note that the $\mathcal Z$ factor only affects
the weak coupling $g_Z$ while the electric charge $e$
attached with photon can be fixed by various electromagnetic
processes and hence should remain the same when discussing
the resonance schemes. Although the pole structure in
the denominators of \geqn{eq:ZA} can be adjusted to be
the same according to \geqn{eq:Zadjust}, the linear
term $s - m^2_{\rm BW}$ and $s - m^2_{\rm ED}$ in
the numerators cannot match simultaneously. In other
words, it is not possible to match both the squared
$|\mathcal M_Z|^2$ and interference $\mathcal M_Z \mathcal M^*_\gamma$
terms at the same time. The difference between the
Breit-Wigner and energy-dependent schemes can arise off
the $Z$ pole with nonvanishing linear terms.

For the total cross section with unpolarized
$e^+ e^-$ beams, both left- and right-handed
electrons can contribute. The $g^2_Z$ term for
the interference term then takes the form as
\cite{Workman:2022ynf},
\begin{align}
  \sigma_{Z \gamma}
=
  \frac {\alpha Q_e Q_f} 6
  \frac {(g_{eL} + g_{eR}) (g_{fL}+g_{fR}) (s - m_Z^2)}
        {(s - m_Z^2)^2 + \Gamma_Z^2 m_Z^2},
\end{align}
for the Breit-Wigner scheme and similarly for other schemes.
The interference term is modulated by $g_{eL} + g_{eR}$ for the
initial $e^+ e^-$ and $g_{fL} + g_{fR}$ for the
final $f \bar f$ fermions. Note that the charged leptons
have $g_{\ell L} = - \frac 1 2 + s^2_w$ and
$g_{\ell R} = s^2_w$ with their prefactor $e/(c_w s_w)$, where $s_w \equiv \sin \theta_w$ ($c_w \equiv \cos \theta_w$)
is the sine (cosine) function of the weak mixing angle $\theta_w$.
With $s^2_w \approx 0.223$, the combination
$g_{\ell L} + g_{\ell R} = - \frac 1 2 + 2 s^2_w$
is quite suppressed. In addition, the interference
term receives another suppression from $s - m^2_Z$.
The interference term is
intrinsically subdominant in the total cross section
of unpolarized $e^+ e^-$ collision.

To make the interference term more explicit,
it is desirable to have either polarized beams
for the initial $e^+ e^-$ or forward-backward
asymmetry for the final-state fermions.
The beam polarization would
replace the electron couplings with
$f_L g_{eL} + f_R g_{eR}$ where $f_L$ and
$f_R$ are the beam polarization fractions
while the final-state factor
$g_{fL} + g_{fR}$ still remains suppressed
for the leptonic final states.
For comparison, the forward-background asymmetry
can fully break the suppression since the cross
section difference \cite{Workman:2022ynf,
ALEPH:2006bhb},
\begin{align}
  \sigma_F - \sigma_B
& \propto
 \frac {3 s}{64 \pi} (g^2_{eL} - g^2_{eR}) (g^2_{fL} - g^2_{fR})
\\
& +
  \frac{3 Q_e Q_f \alpha (s-m^2_Z)} 8
  \left( g_{eL} - g_{eR} \right) \left(g_{fL} - g_{fR}\right),
\nonumber
\end{align}
is intrinsically modulated by $g_{eL} - g_{eR}$
and $g_{fL} - g_{fR}$ both having a magnitude of
$1/2$. Comparing with the
interference term in the second line, the
$|\mathcal M_Z|^2$ term in the first line receives
at least $g_{eL} + g_{eR}$ suppression instead.

As mentioned earlier, it is impossible to match
the Breit-Wigner and energy-dependent schemes in
the presence of both the $Z$ contribution
$|\mathcal M_Z|^2$ and the $Z$-$\gamma$ mediation.
So we take into consideration both the hadronic cross section
$\sigma_{\rm had}$ that mainly provides the
$|\mathcal M_Z|^2$ contribution and the forward-backward asymmetry
$A^f_{\rm FB}$ for the interference term,
\begin{align}
  \sigma_{\rm had} \equiv \sum_q \sigma^q_F +\sigma^q_B,
\quad
  A^f_{FB} \equiv \frac{ \sigma^f_{ F} - \sigma^f_{ B}}{ \sigma^f_{ F} + \sigma^f_{ B} }.
\label{eq:sigmaA}
\end{align}

The previous precision measurements at LEP can already reach
the $10^{-3}$ level \cite{Workman:2022ynf,RothBook}. One may fit
those $Z$ lineshape data from the four detectors
ALEPH \cite{ALEPH:1999smx, ALEPH:2001pzx}, DELPHI \cite{DELPHI:2000wje, DELPHI:2003fml},
L3 \cite{L3:2000vgx, L3:1998bss}, and OPAL \cite{OPAL:2000ufp, OPAL:2003pfe}
at LEP with both the $Z$ mass $m_Z$ and decay width $\Gamma_Z$
as independent parameters. In addition, we also treat
the weak mixing angle $s^2_w$ as an extra fitting
parameters. This fitting can be done for either scheme
to produce its own $\chi^2_{\rm min}$
\begin{table}[h]
\centering
\begin{tabular}{c|cccc|c}
$\Delta \chi^2_{\rm min}$ & ALEPH & DELPHI & L3 & OPAL & Combined  \\
\hline
$\sigma_{\rm had}$ & $0.106$ & $0.0083$ & $0.0127$ & $0.0529$ & $0.183$ \\
$\sigma_{\rm had} + A^\mu_{FB}$&  8.34 & 10.1 & 5.70 & 14.1 & 37.9 \\
$\sigma_{\rm had} + A^{b,c}_{FB}$ &  1.19 & 0.878 & 0.010$^*$ & 0.734 & 2.97 \\
$\sigma_{\rm had} + A^{\mu,b,c}_{FB}$ &  8.42 & 9.67 & 5.58$^*$ & 13.3 & 37.0 \\
\end{tabular}
\caption{The sensitivity of distinguishing the Breit-Wigner
and energy-dependent schemes with the existing LEP data.
The forward-backward asymmetry of a charm quark is not considered at L3 due to its lack of statistics ($^*$).}
\label{tab:LEPchi2min}
\end{table}
and the difference
$\Delta \chi^2_{\rm min} \equiv |\chi^2_{\rm ED, \, min} - \chi^2_{\rm BW, \, min}|$
as summarized in \gtab{tab:LEPchi2min}
characterizes the capability of LEP in distinguishing
the two schemes. Although the $Z$ lineshape
data at various energy points have already been
included, the fit with $\sigma_{\rm had}$ alone does
not have much sensitivity as expected. Only when the $A^f_{FB}$
measurements ($f=\mu,b,c$) are also included, the combined sensitivity
can reach $\Delta \chi^2_{\rm min} = 37$.

At CEPC, a $100 \, {\rm ab}^{-1}$ of integrated luminosity
is projected at the $Z$ pole ($\sqrt{s} \simeq 91.2 \, {\rm GeV}$)
to generate $3 \times 10^{12}$ events \cite{CEPCStudyGroup:2018ghi, CEPCPhysicsStudyGroup:2022uwl}.
The dominant uncertainty
at CEPC comes from the integrated luminosity,
$\delta \mathcal{L} / \mathcal{L} = 5 \times 10^{-5}$
\cite{CEPCPhysicsStudyGroup:2022uwl} 
while its statistical
uncertainty $1/\sqrt{N}$ can reach the $10^{-7}$ level.
Those off-$Z$ pole points
(87.9, 90.2, 92.2, 94.3)\,GeV
would have $1 \, {\rm ab}^{-1}$ each.
Since the forward-backward asymmetry $A^f_{FB}$ in
\geqn{eq:sigmaA} is defined as
ratio of events $N_{F/B} \propto \mathcal L \times \sigma_{F/B}$,
the luminosity $\mathcal L$ and its uncertainty factorize out.
Then the statistical uncertainty,
$(\delta A_{FB})^2 \simeq 4 N_F N_B / (N_F+N_B)^3$,
dominates.

Our fit
$\chi^2 \equiv \sum_i (\mathcal{O}^i_{\rm BW}-\mathcal{O}^i_{\rm ED})^2 / (\delta \mathcal{O}^i)^2$
is done for the hadronic cross
section $\sigma_{\rm had}$, the forward-backward
asymmetries $A^f_{\rm FB}$, and their combination.
The observable pseudo-data $\mathcal{O}^i_{\rm ED}$
and the corresponding uncertainty $\delta \mathcal{O}^i$
are estimated in the energy-dependent scheme with
the PDG values \cite{Workman:2022ynf} of $m_Z = 91.1876$\,GeV,
$\Gamma_Z = 2.4955$\,GeV,
and $s^2_w = 0.223046$. These pseudo-data are then fit
by the Breit-Wigner scheme predictions of $\mathcal{O}^i_{\rm BW}$
with 3 fitting parameters $m_Z$, $\Gamma_Z$, and $s^2_w$.
If either $\sigma_{\rm had}$ or the forward-backward
asymmetries $A^q_{FB}$ ($q=u,d,s,c,b$) are included
in the analysis, there is almost no sensitivity at all.
However, their combination can significantly boost
to $\chi^2_{\rm min} (\sigma_{\rm had}+A^q_{\rm FB}) = 490$.
Further including $A^\mu_{\rm FB}$, $\chi^2_{\rm min}$
jumps to $5.1 \times 10^5$.
Comparing with the LEP result, the $\Delta \chi^2_{\rm min}$
increase by a factor of $1.3 \times 10^4$, being smaller
than the $3 \times 10^5$ times increase in event rates,
is reasonable.

Notice that the soft real and radiative virtual
corrections can
have significant effect on the forward-backward asymmetry
$A^{f}_{\rm FB}$ \cite{Greco:1980mh}. Especially, the
value of $A^{f}_{\rm FB}$ reduces above the $Z$ pole
which would make the relative difference between different
resonance schemes even larger. As a conservative estimation
of the sensitivity, we adopt the tree-level calculation
and leave the more sophisticated soft real and radiative virtual
corrections for future study.

\vspace{2mm}

{\bf Breit-Wigner vs Theoretical Schemes}
--
Between the Breit-Wigner scheme in \geqn{eq:BW} and
the theoretical one in \geqn{eq:th}, the only difference
is that the pole position shifts,
$m^2_{\rm BW} \rightarrow m^2_{\rm th} - \frac 1 4 \Gamma^2_{\rm th}$,
while the imaginary part remains the same,
$m_{\rm BW} \Gamma_{\rm BW} = m_{\rm th} \Gamma_{\rm th}$.
Consequently, the two sets of parameters scale as,
\begin{align}
  \frac {m_{\rm th}}{m_{\rm BW}}
=
  \frac {\Gamma_{\rm BW}}{\Gamma_{\rm th}}
\simeq
  1 + \frac{\Gamma^2_Z}{8 m^2_Z}
\equiv
  R_Z.
\label{eq:mGammaZ}
\end{align}
The relative correction $9.2 \times 10^{-5}$ is smaller
than the difference between the Breit-Wigner and
energy-dependent schemes by a factor of 4.
Since the ratio $\Gamma^2_Z / m^2_Z$ is already a small
number, there is no need to distinguish between
the values of $\Gamma_Z$ and $m_Z$ in the Breit-Wigner
or theoretical schemes. So we omit the scheme labels
for such ratios hereafter.

The equivalence between these two schemes can be
achieved at the amplitude level already, in contrast to
the situation between the Breit-Wigner and
energy-dependent schemes that the exact match
can only happen at the amplitude squared level
$|\mathcal M_Z|^2$.
In other words, the Breit-Wigner and theoretical
schemes share higher similarity. If the $m_Z$
and $\Gamma_Z$ are used as two free parameters,
the Breit-Wigner and theoretical schemes can
fit the $Z$ lineshape equally well \cite{Willenbrock:2022smq}.
Due to the same similarity, the $Z$ lineshape
scan and forward-backward asymmetries discussed
in the previous section can also apply
to tell the difference between the theoretical and
energy-dependent schemes.

The only way for distinguishing the Breit-Wigner
and theoretical schemes probably lies in making
extra connection between $m_Z$ and $\Gamma_Z$.
\geqn{eq:mGammaZ} requires the decay width to be inversely proportional to the mass,
$\Gamma_Z \propto 1 / m_Z$. This relationship arises because the observable quantity as data is the imaginary part of the self-energy function, ${\rm Im} \, \Pi_Z = m_Z \Gamma_Z$.
When extracting the decay width from data,
a heavier $Z$ boson corresponds to a smaller decay width, $\Gamma_Z \propto 1 / m_Z$,
regardless whether the Breit-Wigner or the theoretical
scheme is used. However, the naive expectation with dimensional
analysis tells us that the decay width of an unstable
particle is typically proportional to its mass,
$\Gamma_Z \propto m_Z$ for the $Z \rightarrow f \bar f$
decay,
\begin{align}
  \Gamma_Z
=
  \frac {m_Z}{24 \pi} \beta_f
\left[
  g^2_L + g^2_R
- (g^2_L - 6 g_L g_R + g^2_R)
  \frac {m^2_f}{m^2_Z}
\right]\hspace{-1mm},
\label{eq:GammaZ}
\end{align}
where $m_f$ is the final-state fermion $f$ mass while
$g_L$ and $g_R$ are the left- and right-handed couplings.
To keep formula compact, we have defined the phase space factor
$\beta_f \equiv \sqrt{1 - 4 m^2_f / m^2_Z}$.
The full decay width $\Gamma_Z$ is a total contribution
from quarks, charged leptons, and neutrinos.

This mismatch between the
extracted $\Gamma_Z$ from data and its theoretically
predicted value can be experimentally tested.
To be more specific, once the $Z$ lineshape
fit is done with the Breit-Wigner scheme for
example, the best-fit value of $m_{\rm BW}$ can
be used to predict $\Gamma_Z$ as
\geqn{eq:GammaZ}. Since there is no real data
yet, the central values are taken as the current
$m_Z = 91.1876$\,GeV \cite{Workman:2022ynf} and
the derived $\Gamma_Z = 2.47042$\,GeV with \geqn{eq:GammaZ}.
For such setup, the Breit-Wigner scheme with one
fitting parameter $m_{\rm BW}$ can perfectly
fit the pseudo-data. For the theoretical scheme,
we take $m_{\rm th}$ as the fitting
parameter and $\Gamma_{\rm th}(m_{\rm th})$ is
consequently
predicted as function of $m_{\rm th}$ according
to \geqn{eq:GammaZ}. However the fitting should
still be done after conversion to the Breit-Wigner scheme,
\begin{align}
  \chi^2_Z
\equiv
  \left( \frac {m_{\rm th}/R_Z - m_Z}{\Delta m_Z} \right)^2
+ \left( \frac {\Gamma_{\rm th} R_Z - \Gamma_Z}{\Delta \Gamma_Z} \right)^2,
\end{align}
using the scheme conversion factor $R_Z$ in
\geqn{eq:mGammaZ}.

Comparing with the $9.2 \times 10^{-5}$ correction of
\geqn{eq:mGammaZ}, a thumb rule is that both precisions
$\Delta m_Z \lesssim 8.4$\,MeV and
$\Delta \Gamma_Z \lesssim 0.23$\,MeV are necessary
to have sizable sensitivity.
The current sensitivities
$\Delta m_Z = 2.1$\,MeV and $\Delta \Gamma_Z = 2.3$\,MeV
\cite{Workman:2022ynf} are not good enough.
However, with the projected sensitivity $\Delta m_Z = 0.1$\,MeV
and $\Delta \Gamma_Z = 0.025$\,MeV at CEPC
\cite{CEPCPhysicsStudyGroup:2022uwl} and FCC-ee
\cite{Blondel:2021ema}, $\chi^2_{Z, \rm min}$
can reach 326. For the ILC-GigaZ projection of
$\Delta m_Z = 0.2$\,MeV and $\Delta \Gamma_Z = 0.12$\,MeV
\cite{Belloni:2022due}, the projected sensitivity
can also touch $\chi^2_{Z, \rm min} = 14.3$.
We have also tried implementing the energy-dependent
width for both the Breit-Wigner and theoretical
schemes. Then the $Z$ lineshape scan should be used to
give $\Delta \chi^2_{\rm min} = 349$ and $696$ at CEPC
and FCC-ee, respectively.

\vspace{2mm}

{\bf Fermi Constant and $WW$ Threshold Scan}
--
The Fermi constant $G_F$ has been measured very precisely
in processes such as muon decay \cite{Workman:2022ynf}.
With the Breit-Wigner scheme, the Fermi constant usually
takes the form as,
\begin{align}
  G_{F, \rm BW}
\approx
  \frac {g^2}{4 \sqrt 2 m^2_{W, \rm BW}}
\left(
  1
- \frac {\overline \Gamma^2_W}{2 m^2_W}
\right),
\label{eq:GF-BW}
\end{align}
where $\overline \Gamma_W$ is not necessarily the
full $W$ decay width which will be further elaborated
below. For comparison, the theoretical scheme
in \geqn{eq:th} would lead to,
\begin{align}
  G_{F, \rm th}
\approx
  \frac {g^2}{4 \sqrt 2 m^2_{W, \rm th}}
\left(
  1
- \frac {\overline \Gamma^2_W}{4 m^2_W}
\right).
\end{align}
With the same measurement of the Fermi constant,
the scheme conversion resembles \geqn{eq:mGammaZ}
but with the $W$ mass $m_W$ and reduced decay
with $\overline \Gamma_W$.

However, the $W$ virtuality $p^2$ in the muon decay
process is smaller than the muon mass squared $m^2_\mu$,
$p^2 < m^2_\mu$. The reduced $W$ decay width
$\overline \Gamma_W \approx \frac 2 5 \Gamma_W$
should only consider the partial contributions
\cite{Giacosa:2021mbz} from $W \rightarrow e \nu_e$ and
$u d$. In addition, the energy dependence of the
decay with $\overline \Gamma_W(p^2) = (p^2 / m^2_W)
\overline \Gamma_W$ \cite{Berends:1987bg, Bardin:1988xt, Bardin:1989di}
gives further suppression,
$p^2 / m^2_W \leq m^2_\mu / m^2_W$. Altogether,
the reduced decay width
$\overline \Gamma(p^2) \lesssim 7 \times 10^{-7}
\Gamma_W$ becomes negligibly small. In other words,
the Fermi constant, $G_F \approx g^2 / 4 \sqrt 2 m^2_W$,
would force the
$W$ mass to the same value regardless of the
scheme.

However, a single data input from the Fermi constant
is not enough to distinguish the two interpretations
with the Breit-Wigner and theoretical schemes. It is
necessary to find an independent measurement of the
$W$ mass $m_W$ and decay width $\Gamma_W$,
\begin{align}
\hspace{-2mm}
  \Gamma_W
=
  \frac {m_W}{24 \pi} g^2 \beta_{12}
\left[
  1
- \frac {m^2_1 + m^2_2} {2 m^2_W}
- \frac {(m^2_1 - m^2_2)^2}{2 m^4_W}
\right],
\label{eq:mW}
\end{align}
where the phase space factor is
$\beta_{12} \equiv \beta_+ \beta_-$ with
$\beta_\pm \equiv \sqrt{1 - (m_1 \pm m_2)^2 / m^2_W}$
and $m_{1,2}$ are the final-state fermion masses.
Considering the measured sizes, $m_W = 80.377$\,GeV
and $\Gamma_W = 2.085$\,GeV \cite{Workman:2022ynf},
the correction $8.1 \times 10^{-5}$
is larger than the Fermi constant precision
\cite{Workman:2022ynf}.

At the future lepton colliders,
the $WW$ threshold scan can provide precise measurement
of $m_W$ and $\Gamma_W$. For example, the projected
precision can reach $\Delta m_W = 0.5$\,MeV and
$\Delta \Gamma_W = 2.0$\,MeV at CEPC
\cite{CEPCPhysicsStudyGroup:2022uwl},
$\Delta m_W = 0.4$\,MeV and
$\Delta \Gamma_W = 1.2$\,MeV at FCC \cite{Blondel:2021ema},
as well as $\Delta m_W = 2.4$\,MeV and
$\Delta \Gamma_W = 2$\,MeV at ILC \cite{deBlas:2022ofj}.
Following the same procedure as the $Z$
mass and decay width fit in the previous
section, a single-parameter fit can also
be done for the $WW$ threshold scan
that provides a simultaneous determination
of $m_W$ and $\Gamma_W$, in addition to
the Fermi constant,
\begin{align}
  \chi^2_W
& \equiv
  \left( \frac {m_{W, \rm th}/R_W - m_W}{\Delta m_W} \right)^2
+ \left( \frac {\Gamma_{W, \rm th} R_W - \Gamma_W}{\Delta \Gamma_W} \right)^2
\nonumber
\\
& +
  \left( \frac {G_{F, \rm th} - G_F}{\Delta G_F} \right)^2,
\end{align}
where $R_W$ takes the similar form as $R_Z$
in \geqn{eq:mGammaZ}.
For the central values we take the current
$m_W = 80.376$\,GeV \cite{Workman:2022ynf}
and the derived $\Gamma_W = 2.04501$ from
\geqn{eq:mW}, in addition to
$G_F = 1.166\,378\,8 \times 10^{-5}\,\mbox{GeV}^{-2}$
\cite{Workman:2022ynf}. The weak coupling
is then fixed to $g = 0.652986$ by \geqn{eq:GF-BW}
in the Breit-Wigner scheme. The sensitivity
can reach $\Delta \chi^2_{W, \rm min} = 169$,
263, and 7.3 at CEPC, FCC-ee, and ILC,
respectively. By fitting the $WW$ threshold
scan data, the extracted $W$ mass values from
the Breit-Wigner and theoretical schemes should
also follow the scheme conversion in \geqn{eq:mGammaZ}.
However, the precise measurement of the Fermi constant
tends to give exactly the same value for $m_W$
in both schemes. These different scaling behaviors provide the
distinguishing power.

\vspace{2mm}

{\bf Conclusions}
--
Although various resonance schemes of unstable
particles have been extensively studied, using
electroweak precision measurement data to verify
which scheme is correct is still an open
question. With simple phenomenological $\chi^2$
fit, we found that the combination of
$Z$ lineshape scan and the forward-backward
asymmetry measurements at the $Z$ pole can
tell the difference between the energy-dependent
scheme and its Breit-Wigner (theoretical)
counterpart with the existing data from LEP.
Even higher sensitivity can be expected at the
future CEPC and FCC-ee. For the remaining Breit-Wigner
and theoretical schemes, they were thought to be
fully equivalent and give exactly the same resonance
shape, we also found a way of using the
opposite scaling behaviors of the extracted
decay width from data, $\Gamma_Z \propto 1 / m_Z$,
and the theoretically predicted one,
$\Gamma_Z \propto m_Z$, to set a realistic approach
for future lepton colliders. For the $WW$
threshold scan, the supplementation from the
Fermi constant is necessary. More detailed simulation
and data analysis from the experimental side is necessary.
In summary, the resonance
shape alone cannot distinguish its various
schemes but needs to involve the forward-backward
asymmetries and the scaling behaviors of decay
width. These provide realistic tests of an
important aspect of quantum and field theories.

\vspace{2mm}

\section*{Acknowledgements}
The authors are grateful to Manqi Ruan and
Junping Tian for providing useful information.
SFG and UM are supported by the National Natural Science
Foundation of China (12425506, 12375101, 12090060 and 12090064) and the SJTU Double First
Class start-up fund (WF220442604).
ZQ is supported by the National Natural Science
Foundation of China (1240050404) and the HZNU start-up fund.
SFG is also an affiliate member of Kavli IPMU, University of Tokyo.


\addcontentsline{toc}{section}{References}
\begin{thebibliography}{99}

\bibitem{Breit:1936zzb}
  G.~Breit and E.~Wigner,
  ``{\it Capture of Slow Neutrons},''
  \href{http://dx.doi.org/10.1103/PhysRev.49.519}
  {Phys. Rev. \textbf{49}, 519-531 (1936)}.

\bibitem{Bohm:2004zi}
  A.~R.~Bohm and Y.~Sato,
  ``{\it Relativistic resonances: Their masses, widths, lifetimes, superposition, and causal evolution},''
  \href{http://dx.doi.org/10.1103/PhysRevD.71.085018}
  {Phys. Rev. D \textbf{71}, 085018 (2005)}
  [\href{http://arxiv.org/abs/hep-ph/0412106}{arXiv:hep-ph/0412106} [hep-ph]].

\bibitem{Giacosa:2021mbz}
  F.~Giacosa, A.~Okopi\'nska and V.~Shastry,
  ``{\it A simple alternative to the relativistic Breit\textendash{}Wigner distribution},''
  \href{http://dx.doi.org/10.1140/epja/s10050-021-00641-2}
  {Eur. Phys. J. A \textbf{57}, no.12, 336 (2021)}
  [\href{http://arxiv.org/abs/2106.03749}{arXiv:2106.03749} [hep-ph]].
   
\bibitem{Dalitz:1961dx}
 R.~H.~Dalitz,
 ``{\it On the strong interactions of the strange particles},''
 \href{http://dx.doi.org/10.1103/RevModPhys.33.471}
 {Rev. Mod. Phys. \textbf{33}, 471-492 (1961)}

\bibitem{Dalitz:1963xk}
R.~H.~Dalitz,
 ``{\it Strange-particle resonant states},''
 \href{http://dx.doi.org/10.1146/annurev.ns.13.120163.002011}
 {Ann. Rev. Nucl. Part. Sci. \textbf{13}, 339-430 (1963)}

\bibitem{Willenbrock:1991hu}
  S.~Willenbrock and G.~Valencia,
  ``{\it On the definition of the Z boson mass},''
  \href{http://dx.doi.org/10.1016/0370-2693(91)90843-F}
  {Phys. Lett. B \textbf{259}, 373-376 (1991)}.

\bibitem{Willenbrock:2022smq}
  S.~Willenbrock,
  ``{\it Mass and width of an unstable particle},''
  \href{http://dx.doi.org/10.1140/epjp/s13360-024-05301-0}
  {Eur. Phys. J. Plus \textbf{139}, no.6, 523 (2024)}
  [\href{http://arxiv.org/abs/2203.11056}{arXiv:2203.11056} [hep-ph]].

\bibitem{Sirlin:1991fd}
  A.~Sirlin,
  ``{\it Theoretical considerations concerning the Z0 mass},''
  \href{http://dx.doi.org/10.1103/PhysRevLett.67.2127}
  {Phys. Rev. Lett. \textbf{67}, 2127-2130 (1991)}.

\bibitem{Papavassiliou:1995fq}
  J.~Papavassiliou and A.~Pilaftsis,
  ``{\it Gauge invariance and unstable particles},''
  \href{http://dx.doi.org/10.1103/PhysRevLett.75.3060}
  {Phys. Rev. Lett. \textbf{75}, 3060-3063 (1995)}
  [\href{http://arxiv.org/abs/hep-ph/9506417}{arXiv:hep-ph/9506417} [hep-ph]];
%
  J.~Papavassiliou and A.~Pilaftsis,
  ``{\it A Gauge independent approach to resonant transition amplitudes},''
  \href{http://dx.doi.org/10.1103/PhysRevD.53.2128}
  {Phys. Rev. D \textbf{53}, 2128-2149 (1996)}
  [\href{http://arxiv.org/abs/hep-ph/9507246}{arXiv:hep-ph/9507246} [hep-ph]];
%
  J.~Papavassiliou and A.~Pilaftsis,
  ``{\it Gauge invariant resummation formalism for two point correlation functions},''
  \href{http://dx.doi.org/10.1103/PhysRevD.54.5315}
  {Phys. Rev. D \textbf{54}, 5315-5335 (1996)}
  [\href{http://arxiv.org/abs/hep-ph/9605385}{arXiv:hep-ph/9605385} [hep-ph]].

\bibitem{Workman:2022ynf}
  S.~Navas \textit{et al.} [Particle Data Group],
  ``{\it Review of Particle Physics},''
  \href{https://doi.org/10.1103/PhysRevD.110.030001}
  {Phys. Rev. D \textbf{110}, 030001 (2024)}.

\bibitem{Berends:1987bg}
  F.~A.~Berends, G.~Burgers, W.~Hollik and W.~L.~van Neerven,
  ``{\it The Standard $Z$ Peak},''
  \href{http://dx.doi.org/10.1016/0370-2693(88)91593-6}
  {Phys. Lett. B \textbf{203}, 177-182 (1988)}
  (CERN-TH 4919/87).

\bibitem{Bardin:1988xt}
  D.~Y.~Bardin, A.~Leike, T.~Riemann and M.~Sachwitz,
  ``{\it Energy Dependent Width Effects in e+ e- Annihilation Near the Z Boson Pole},''
  \href{http://dx.doi.org/10.1016/0370-2693(88)91627-9}
  {Phys. Lett. B \textbf{206}, 539-542 (1988)}.

\bibitem{Bardin:1989di}
  D.~Y.~Bardin, M.~S.~Bilenky, G.~Mitselmakher, T.~Riemann and M.~Sachwitz,
  ``{\it A Realistic Approach to the Standard Z Peak},''
  \href{http://dx.doi.org/10.1007/BF01415565}
  {Z. Phys. C \textbf{44}, 493 (1989)}.

\bibitem{WorkingGrouponLEPEnergy:1993sza}
  L.~Arnaudon \textit{et al.} [Working Group on LEP Energy, ALEPH, DELPHI, L3 and OPAL],
  ``{\it Measurement of the mass of the Z boson and the energy calibration of LEP},''
  \href{http://dx.doi.org/10.1016/0370-2693(93)90210-9}
  {Phys. Lett. B \textbf{307}, 187-193 (1993)}.

\bibitem{ALEPH:2005ab}
  S.~Schael \textit{et al.} [ALEPH, DELPHI, L3, OPAL, SLD, LEP Electroweak Working Group, SLD Electroweak Group and SLD Heavy Flavour Group],
  ``{\it Precision electroweak measurements on the $Z$ resonance},''  
  \href{http://dx.doi.org/10.1016/j.physrep.2005.12.006}
  {Phys. Rept. \textbf{427}, 257-454 (2006)}
  [\href{https://arxiv.org/pdf/hep-ex/0509008}{arXiv:hep-ex/0509008} [hep-ex]].

\bibitem{Bardin:1999yd}
  D.~Y.~Bardin, P.~Christova, M.~Jack, L.~Kalinovskaya, A.~Olchevski, S.~Riemann and T.~Riemann,
  ``{\it ZFITTER v.6.21: A Semianalytical program for fermion pair production in $e^+ e^-$ annihilation},''
  \href{http://dx.doi.org/10.1016/S0010-4655(00)00152-1}
  {Comput. Phys. Commun. \textbf{133}, 229-395 (2001)}
  [\href{http://arxiv.org/abs/hep-ph/9908433}{arXiv:hep-ph/9908433} [hep-ph]].

\bibitem{ALEPH:2013dgf}
S.~Schael \textit{et al.} [ALEPH, DELPHI, L3, OPAL and LEP Electroweak],
``{\it Electroweak Measurements in Electron-Positron Collisions at W-Boson-Pair Energies at LEP},''
 \href{http://dx.doi.org/10.1016/j.physrep.2013.07.004}
 {Phys. Rept. \textbf{532}, 119-244 (2013)}
 [\href{http://arxiv.org/abs/1302.3415}{arXiv:1302.3415} [hep-ex]].

\bibitem{ATLAS:2024erm}
G.~Aad \textit{et al.} [ATLAS],
``{\it Measurement of the W-boson mass and width with the ATLAS detector using proton\textendash{}proton collisions at $\sqrt{s}=7$ TeV},''
\href{http://dx.doi.org/10.1140/epjc/s10052-024-13190-x}
{Eur. Phys. J. C \textbf{84}, no.12, 1309 (2024)}
[\href{http://arxiv.org/abs/2403.15085}{arXiv:2403.15085} [hep-ex]].

\bibitem{CMS:2024nau}
 [CMS],
 ``{\it Measurement of the W boson mass in proton-proton collisions at $\sqrt{s} = 13$\,TeV},''
 \href{https://cms-results.web.cern.ch/cms-results/public-results/preliminary-results/SMP-23-002}
 {CMS-PAS-SMP-23-002}.

\bibitem{CEPCPhysicsStudyGroup:2022uwl}
  H.~Cheng \textit{et al.} [CEPC Physics Study Group],
  ``{\it The Physics potential of the CEPC. Prepared for the US Snowmass Community Planning Exercise (Snowmass 2021)},''
  [\href{http://arxiv.org/abs/2205.08553}{arXiv:2205.08553} [hep-ph]].

\bibitem{Blondel:2021ema}
  A.~Blondel and P.~Janot,
  ``{\it FCC-ee overview: new opportunities create new challenges},''
  \href{http://dx.doi.org/10.1140/epjp/s13360-021-02154-9}
  {Eur. Phys. J. Plus \textbf{137}, no.1, 92 (2022)}
  [\href{http://arxiv.org/abs/2106.13885}{arXiv:2106.13885} [hep-ex]].

\bibitem{ALEPH:2006bhb}
  J.~Alcaraz \textit{et al.} [ALEPH, DELPHI, L3, OPAL and LEP Electroweak Working Group],
  ``{\it A Combination of preliminary electroweak measurements and constraints on the standard model},''
  [\href{http://arxiv.org/abs/hep-ex/0612034}{arXiv:hep-ex/0612034} [hep-ex]].

\bibitem{RothBook}
  Stefan Roth,
  ``{\it Precision electroweak physics at electron-positron colliders},''
  \href{https://doi.org/10.1007/3-540-35165-5}
  {Springer Tracts in Modern Physics Vol 220,
  Sprinter 2007},
  ISBN-10 3-540-35164-7,
  ISBN-13 978-3-540-35164-1.

\bibitem{ALEPH:1999smx}
  R.~Barate \textit{et al.} [ALEPH],
  ``{\it Measurement of the Z resonance parameters at LEP},''
  \href{http://dx.doi.org/10.1007/s100520000319}
  {Eur. Phys. J. C \textbf{14}, 1-50 (2000)}.

\bibitem{ALEPH:2001pzx}
  A.~Heister \textit{et al.} [ALEPH],
  ``{\it Measurement of the forward backward asymmetry in $Z \rightarrow b \bar b$ and $Z \rightarrow c \bar c$ decays with leptons},''
  \href{http://dx.doi.org/10.1007/s100520200950}
 {Eur. Phys. J. C \textbf{24}, 177-191 (2002)}.

\bibitem{DELPHI:2000wje}
  P.~Abreu \textit{et al.} [DELPHI],
  ``{\it Cross-sections and leptonic forward backward asymmetries from the Z0 running of LEP},''
  \href{http://dx.doi.org/10.1007/s100520000392}
  {Eur. Phys. J. C \textbf{16}, 371-405 (2000)}.

\bibitem{DELPHI:2003fml}
  J.~Abdallah \textit{et al.} [DELPHI],
  ``{\it Measurement of the forward backward asymmetries of $e^+ e^- \rightarrow Z \rightarrow b \bar b$ and $e^+ e^- \rightarrow Z \rightarrow c \bar c$ using prompt leptons},''
  \href{http://dx.doi.org/10.1140/epjc/s2004-01708-6}
  {Eur. Phys. J. C \textbf{34}, 109-125 (2004)}
  [\href{http://arxiv.org/abs/hep-ex/0403041} {arXiv:hep-ex/0403041} [hep-ex]].

\bibitem{L3:2000vgx}
  M.~Acciarri \textit{et al.} [L3],
  ``{\it Measurements of cross-sections and forward backward asymmetries at the $Z$ resonance and determination of electroweak parameters},''
  \href{http://dx.doi.org/10.1007/s100520050001}
  {Eur. Phys. J. C \textbf{16}, 1-40 (2000)}
  [\href{http://arxiv.org/abs/hep-ex/0002046}{arXiv:hep-ex/0002046} [hep-ex]].

\bibitem{L3:1998bss}
  M.~Acciarri \textit{et al.} [L3],
  ``{\it Measurement of the $e^+ e^- \rightarrow Z \rightarrow b \bar b$ forward-backward asymmetry and the B0 anti-B0 mixing parameter using prompt leptons},''
  \href{http://dx.doi.org/10.1016/S0370-2693(98)01601-3}
  {Phys. Lett. B \textbf{448}, 152-162 (1999)}. 

\bibitem{OPAL:2000ufp}
  G.~Abbiendi \textit{et al.} [OPAL],
  ``{\it Precise determination of the Z resonance parameters at LEP: 'Zedometry'},''
  \href{http://dx.doi.org/10.1007/s100520100627}
  {Eur. Phys. J. C \textbf{19}, 587-651 (2001)}
  [\href{http://arxiv.org/abs/hep-ex/0012018}{arXiv:hep-ex/0012018} [hep-ex]].

\bibitem{OPAL:2003pfe}
G.~Abbiendi \textit{et al.} [OPAL],
``{\it Measurement of heavy quark forward backward asymmetries and average B mixing using leptons in hadronic Z decays},''
  \href{http://dx.doi.org/10.1016/j.physletb.2003.10.022}
{Phys. Lett. B \textbf{577}, 18-36 (2003)}
  [\href{http://arxiv.org/abs/hep-ex/0308051}{arXiv:hep-ex/0308051} [hep-ex]].  

\bibitem{CEPCStudyGroup:2018ghi}
  J.~B.~Guimar\~aes da Costa \textit{et al.} [CEPC Study Group],
  ``{\it CEPC Conceptual Design Report: Volume 2 - Physics \& Detector},''
  [\href{http://arxiv.org/abs/1811.10545}{arXiv:1811.10545} [hep-ex]].

\bibitem{Greco:1980mh}
M.~Greco, G.~Pancheri-Srivastava and Y.~Srivastava,
``{\it Radiative Corrections to e+ e- ---\ensuremath{>} mu+ mu- Around the Z0},''
\href{https://doi.org/10.1016/0550-3213(80)90363-6}
{Nucl. Phys. B \textbf{171}, 118 (1980)}
  
\bibitem{Belloni:2022due}
  A.~Belloni, A.~Freitas, J.~Tian, J.~Alcaraz Maestre, A.~Apyan, B.~Azartash-Namin, P.~Azzurri, S.~Banerjee, J.~Beyer and S.~Bhattacharya, \textit{et al.}
  ``{\it Report of the Topical Group on Electroweak Precision Physics and Constraining New Physics for Snowmass 2021},''
  [\href{http://arxiv.org/abs/2209.08078}{arXiv:2209.08078} [hep-ph]].

\bibitem{deBlas:2022ofj}
  J.~de Blas, Y.~Du, C.~Grojean, J.~Gu, V.~Miralles, M.~E.~Peskin, J.~Tian, M.~Vos and E.~Vryonidou,
  ``{\it Global SMEFT Fits at Future Colliders},''
  [\href{http://arxiv.org/abs/2206.08326}{arXiv:2206.08326} [hep-ph]].

\end{thebibliography}

\end{document}